\documentclass{article}
\pdfoutput=1
\usepackage{graphicx}
\usepackage{epsfig}
\usepackage{amssymb}
\usepackage{amsmath}
\usepackage{upgreek,bm}
\usepackage{fullpage}

\title{Natural convection in eccentric spherical annuli} 

\author{A. D. Gallegos \footnotemark[2] 
\and C. M\'alaga \footnotemark[2] \thanks{Author to whom correspondence should be addressed:
cmi.ciencias@ciencias.unam.mx}
}

\date{}

\begin{document}
\maketitle

\renewcommand{\thefootnote}{\fnsymbol{footnote}}
\footnotetext[2]{Physics Department, School of Science, Universidad Nacional Aut\'onoma de M\'exico}

\renewcommand{\thefootnote}{\arabic{footnote}}

\begin{abstract}
A fluid between two spheres, concentric or not, at different temperatures will flow in the presence of a constant
gravitational force. Although there is no possible hydrostatic state, energy transport is dominated by diffusion
if temperature difference between the spheres is small enough. In this conductive regime the average Nusselt 
number remains approximately constant for all Rayleigh numbers below some critical value. Above the critical 
Rayleigh number, plumes appear and thermal convection takes place. We study this phenomenon, in particular 
the case where the inner sphere is displaced from the centre, using a two-component thermal lattice Boltzmann 
method to characterize the convective instability, the evolution of the flow patterns and the dependence of the 
Nusselt number on the Rayleigh number beyond the transition.

\end{abstract}


\section{Introduction}

Natural convective flow between a sphere at constant temperature and its spherical enclosure at a different 
temperature is an idealisation of many problems of practical interest. Convection patterns between concentric 
spherical shells were first observed by Bishop {\it et al.} in 1966 
for a range of aspect ratios \cite{Bishop1966}. Experiments were performed with the inner sphere hotter than the 
outer sphere. A wide variety of steady and non steady patterns were observed in subsequent 
experiments using air, water and silicone oil; including the case of vertically eccentric spherical shells 
\cite{Powe1980}. For concentric spheres, analytic solutions for the conductive regime were reported by 
Makc \& Hardee \cite{Lawrence1968}. The 
dependence of the Nusselt number $Nu$ on the Rayleigh number $Ra$ on the transition from the conductive to 
the convective steady solutions was obtained by Teertstra {\it et al.} \cite{Teertstra2006}. The stability analysis of 
the convective regime has deserved wide 
attention, see for example \cite{Travnikov2015}. Numerical simulations of the concentric configuration can be 
found for a variety of numerical methods, in particular for the study of the transition to turbulence 
\cite{Colonius2013, Scurtu2010, Chu1993, Dehghan2010, Chiu1996}.

Little is known of the inverted configuration. The case of an outer shell hotter than an inner concentric shell was 
experimentally and numerically studied by 
Futterer {\it et al.} using silicone oil \cite{Futterer2007}. They found unsteady periodic flows consisting of cold 
blobs dripping from the inner sphere at a frequency that became irregular through a period-doubling process as 
$Ra$ increases.     

While concentric configurations have received much attention, few studies can be found about the eccentric 
ones. In the present work, we report a numerical exploration of the natural convection between eccentric 
spherical shells at fixed temperatures (see figure \ref{fig1}). Simulations were performed using a three 
dimensional, two-component lattice 
Boltzmann method (LBM) that approximates solutions to the Oberbeck-Boussinesq equations of thermal 
convection \cite{Inamuro}. This method has been used successfully for the simulation of natural convection 
phenomena \cite{Mandujano,Mehrizi2013}.

In the next section we present the numerical method and its validation. We compare predicted results with 
concentric and vertically eccentric numerical and experimental observations found in the literature. We present 
numerical solutions for different eccentric configurations in the third section. The transition from conductive to
convective heat transport is characterised, and a series of steady and unsteady patterns and behaviour are 
presented in this section. In the last section we summarise our observations and motivate future work.

\section{The thermal LBM}

To simulate natural convection phenomena we used a D3Q19 two-component lattice Boltzmann equation. The 
method is a direct three dimensional extension of that proposed by Innamuro in 2002 \cite{Inamuro}. Space is 
discretized using a cubic lattice where a density $f_{k}$ and a temperature ${\text g}_{k}$ distribution functions are 
computed. The distribution functions are then used to compute the fluid velocity ${\boldsymbol u}$ and 
temperature $T$ at the lattice nodes. Lattice spacing as well as time steps can be conveniently set to unity. At 
every node $\boldsymbol r$ in the lattice, the distribution functions evolve in time 
according to
\begin{eqnarray}
 f_{k} ({\boldsymbol r} + {\boldsymbol e}_{k}, t + 1) & = & 
 f_{k} ({\boldsymbol r}, t) - \tfrac{1}{\tau} \left[ f_{k} ({\boldsymbol r}, t)
 - f_{k}^{eq} ({\boldsymbol r}, t) \right] + G_{k},
 \label{lbmf}  \\
 g_{k} ({\boldsymbol r} + {\boldsymbol e}_{k}, t + 1 ) & = & 
 g_{k} ({\boldsymbol r}, t) - \tfrac{1}{\tau_g} \left[ g_{k} ({\boldsymbol r}, t)
 - g_{k}^{eq} ({\boldsymbol r}, t) \right].
\label{lbmg}
\end{eqnarray}

The coefficients $\tau$ and $\tau_g$ represent relaxation times and are related to the fluid kinematic viscosity 
$\nu = (\tau - 1/2)/3$ and thermal diffusivity $\alpha = (\tau_g - 1/2)/3$. The local equilibrium distribution 
functions $f_{k}^{eq}$ and $g_{k}^{eq}$ are given by
\begin{eqnarray}
 f_{k}^{eq} & = & \rho w_k \left[ 1 + 3 {\boldsymbol e}_{k}  {\boldsymbol \cdot} {\boldsymbol u} + \tfrac{9}{2} \left(
 {\boldsymbol e}_{k} {\boldsymbol \cdot} {\boldsymbol u} \right)^2 -  \tfrac{3}{2} u^2 \right], 
 \label{eqf} \\
 g_{k}^{eq} & = & T w_k \left[ 1 + 3 {\boldsymbol e}_{k} {\boldsymbol \cdot} {\boldsymbol u} \right].
 \label{eqg}
\end{eqnarray}

The equilibrium distributions depend on the macroscopic fields ${\boldsymbol u}$, $T$ and $\rho$, the mass density, and must be computed every time step through
 \begin{eqnarray}
 \rho ({\boldsymbol r}, t) & = & \sum_{k=0}^{18}  f_{k}({\boldsymbol r}, t), \\
 \rho  {\boldsymbol u}({\boldsymbol r}, t) & = & \sum_{k=0}^{18}  {\boldsymbol e}_{k} f_{k}({\boldsymbol r}, t), \\
 T ({\boldsymbol r}, t) & = & \sum_{k=0}^{18}  g_{k}({\boldsymbol r}, t)
 \label{macros}
\end{eqnarray}

The constants $w_k$ in the equilibrium functions definitions (\ref{eqf}) and  (\ref{eqg}) take the values $w_0 = 
1/3$, $w_k = 1/18$ for 
$k = 1,...,6$ and  $w_k = 1/36$ for $k = 7,...,18$. The last term in (\ref{lbmf}) is related to the body
force and gives the buoyancy term in the Boussinesq equations. It is defined as  $G_k = -3 \beta w_k 
\left( T({\boldsymbol r}, t) - T_0 \right) {\boldsymbol e}_{k}  {\boldsymbol \cdot} {\boldsymbol g}$, where $\beta$ is the coefficient 
of thermal expansion of the fluid, ${\boldsymbol g}$ the acceleration due to gravity and $T_0$ a reference temperature 
taken as the average of the temperatures of the inner and the outer shells.

The set of microscopic velocities $\left\{ {\boldsymbol e}_{k}: \ \ k=0,...,18 \right\}$ is given by
\begin{eqnarray}
 {\boldsymbol e}_0 = (0,0,0), \ \ \ \ {\boldsymbol e}_1 = - {\boldsymbol e}_4 =(1,0,0), 
\ \ \ \ {\boldsymbol e}_2 = - {\boldsymbol e}_5 =(0,1,0), \ \ \ \ {\boldsymbol e}_3 = - {\boldsymbol e}_6 =(0,0,1), 
\nonumber \\
{\boldsymbol e}_7 = - {\boldsymbol e}_{10} =(1,1,0), 
\ \ \ \ {\boldsymbol e}_8= - {\boldsymbol e}_{11} =(1,0,1), \ \ \ \ {\boldsymbol e}_9 = - {\boldsymbol e}_{12} 
=(0,1,1), 
\nonumber \\
{\boldsymbol e}_{13} = - {\boldsymbol e}_{16} =(-1,1,0), \ \ \ \ {\boldsymbol e}_{14}= - {\boldsymbol e}_{17} 
=(-1,0,1), \ \ \ \ {\boldsymbol e}_{15} = - {\boldsymbol e}_{18} =(0,-1,1).
\end{eqnarray} 

Equations (\ref{lbmf}) and (\ref{lbmg}), with the choice microscopic velocities, provide an algorithm for 
updating all 
the distribution functions $f_k$ and $g_k$ at a given node in the lattice, as long as its $18$ nearest neighbours in 
the lattice are inside the fluid domain. For nodes close enough to a solid wall, the distribution functions coming 
from neighbouring nodes outside the fluid domain must be provided as a boundary condition for the method. We 
choose to adopt the set of boundary conditions proposed by  \cite{Mehrizi2013} for curved solid walls at fixed 
temperatures. In this boundary conditions, distribution functions that should provide the node outside the fluid domain are extrapolated 
considering that $T$ and  ${\boldsymbol u}$ are prescribed at the point in the boundary that intersects the line 
joining the nodes in question.

The method was implemented to run in parallel in Graphic Processing Units (GPU) because of the large number 
of nodes needed in the simulation of three dimensional flows. Typically, around $10^7$ nodes where used to 
obtain unsteady flows and simulations took a few days running on a single GPU.

\section{Problem statement and validation}

\begin{figure}[t]
\centering
\setlength{\unitlength}{0.1\textwidth}
  \begin{picture}(7,2.5)
    \put(0.5,0){\includegraphics[width=0.25\columnwidth]{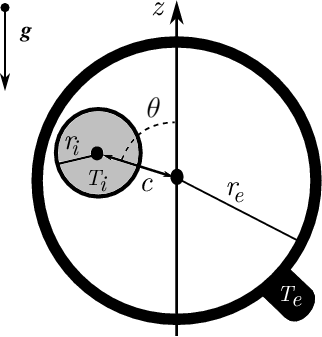}}
    \put(4,0){\includegraphics[width=0.25\columnwidth]{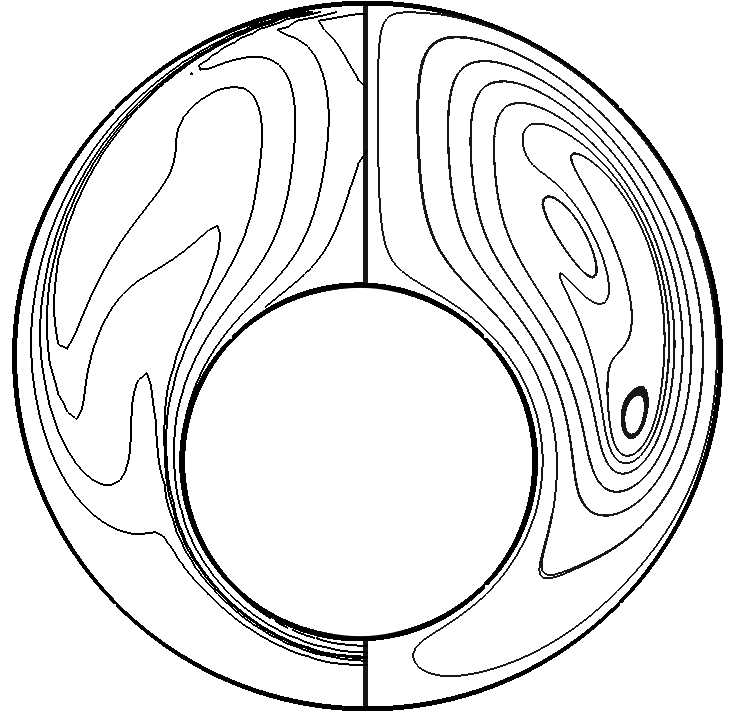}}
    \put(0,2.4){(a)}
    \put(3.8,2.4){(b)}
  \end{picture}
\caption{ (a) Schematic representation of the problem. (b) Steady numerical solution for $\eta= 0.5$, $\theta = 
 \pi$, $\epsilon = 0.6$ and $Ra = 1.875 \times 10^5$. The numerical solutions is symmetric with respect to the 
 vertical axis and the figure shows isothermal curves in the left half  of a 
 plane of symmetry and stream lines in the right half.}
\label{fig1}
\end{figure}

A fluid of initial density $\rho_0$ fills the gap between two eccentric spherical shells at fixed positions. The inner sphere of radius $r_i$ is at
temperature $T_i$ and the external sphere of radius $r_e$ is at a higher temperature $T_e$. The position of the centre of 
the inner sphere is given by the polar angle $\theta$ and the distance to the external sphere centre $c$, as 
shown in figure \ref{fig1}(a). The vertical plain containing the centres of both spheres is a plane of symmetry of 
the fluid domain.

The thermal LBM presented in the previous section approximates solutions to the Boussinesq equations 
describing the convective flow of the stated problem \cite{Inamuro}. Scaling lengths with $L=r_e - r_i$, 
temperatures with 
$\Delta T =  T_i - T_e$, velocities with $\alpha / L$ and pressure with $\rho_0 \alpha^2 / L^2$, the 
convective flow is defined by the non-dimensional eccentricity $\epsilon$, the polar angle $\theta$, 
the aspect ratio $\eta$, the Rayleigh number $Ra$ and the Prandtl number $Pr$ given by
\begin{equation}
\epsilon = \frac{c}{r_e}, \ \ \ \ \eta= \frac{r_e}{r_i},  \ \ \ \ Ra=\frac{g \beta \Delta T L^3}{\nu \alpha}, 
 \ \ \ \ Pr=\frac{\nu}{\alpha}.
\end{equation}
For all the flows studied $Pr = 0.7$, corresponding to air, and $T_i > T_e$ which gives $Ra >0$.

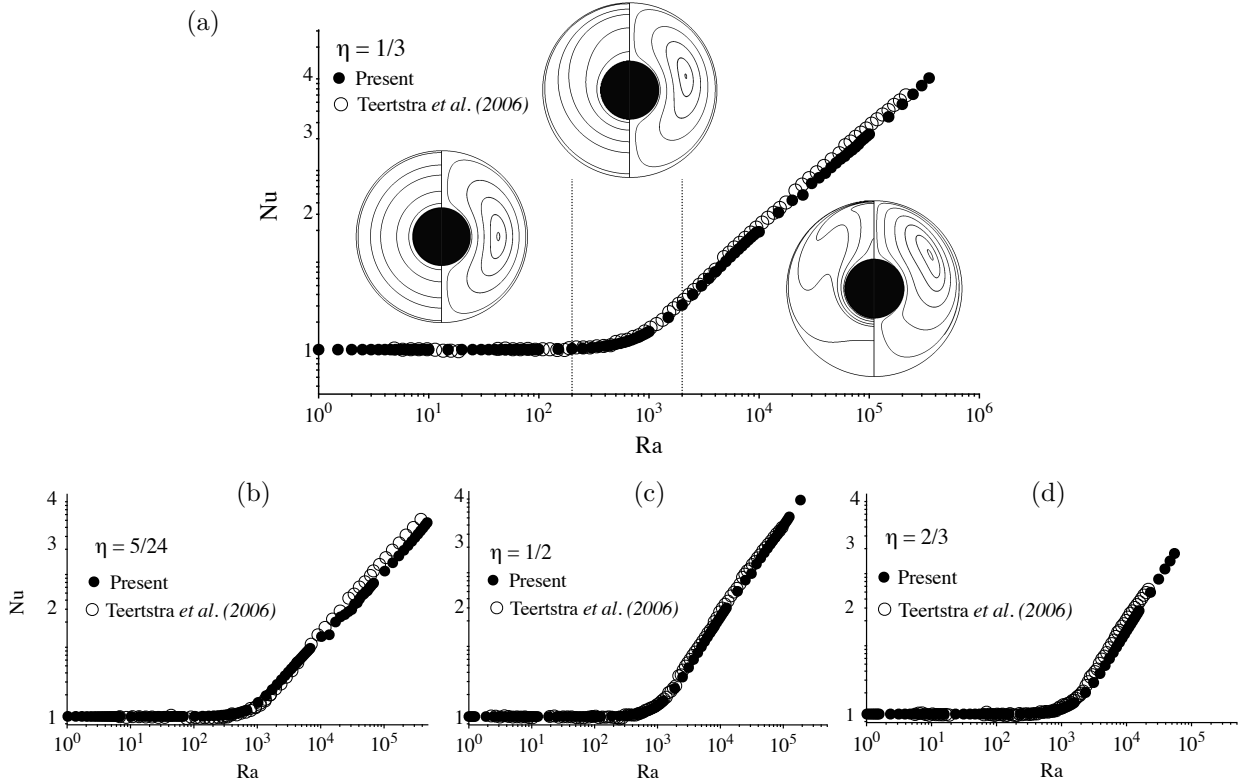
\begin{figure}[t]
\centering
\setlength{\unitlength}{0.1\textwidth}
  \begin{picture}(10,6.3)
    \put(2,2.6){\includegraphics[width=0.6\columnwidth]{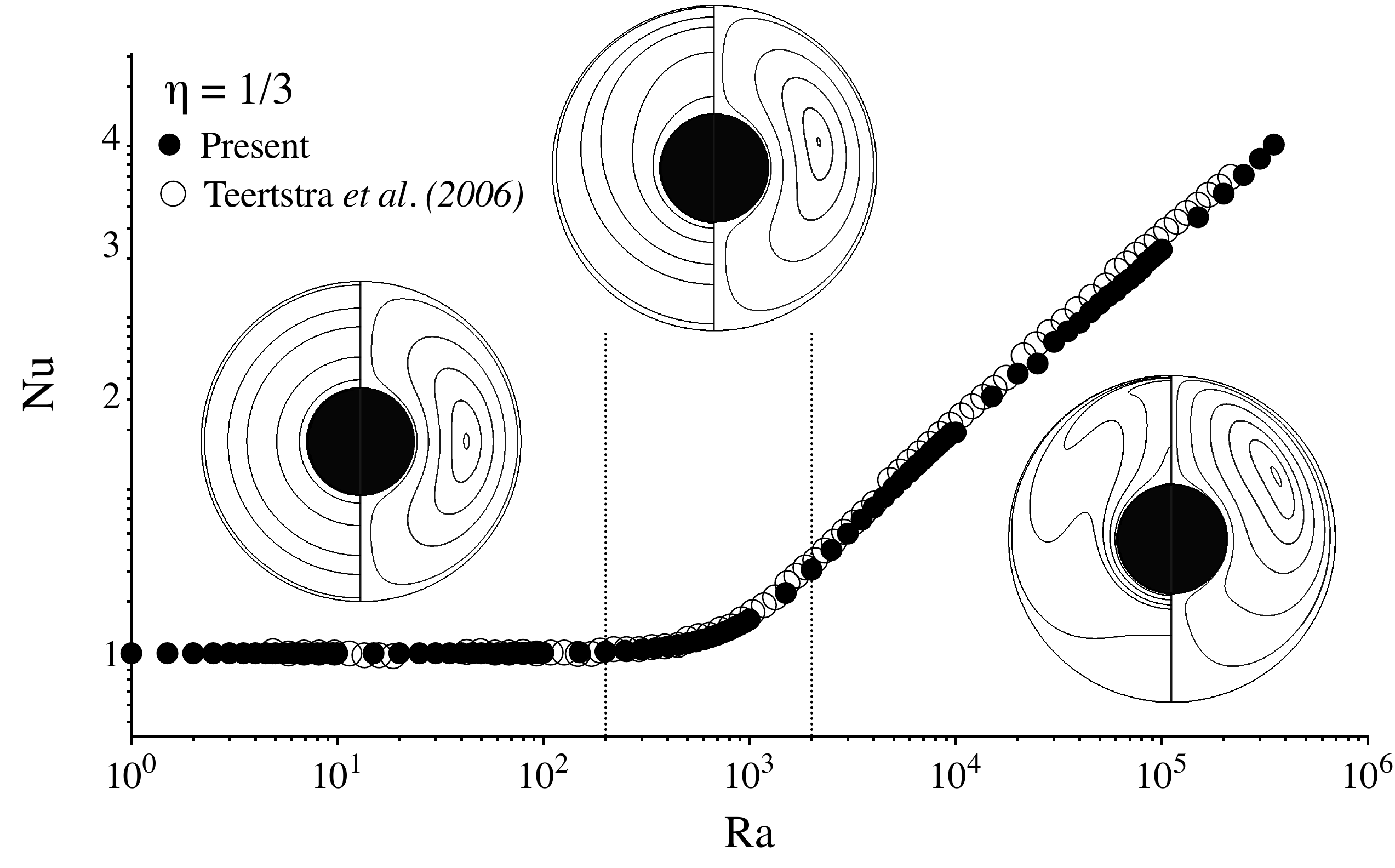}}
    \put(6.5,0){\includegraphics[width=0.35\columnwidth]{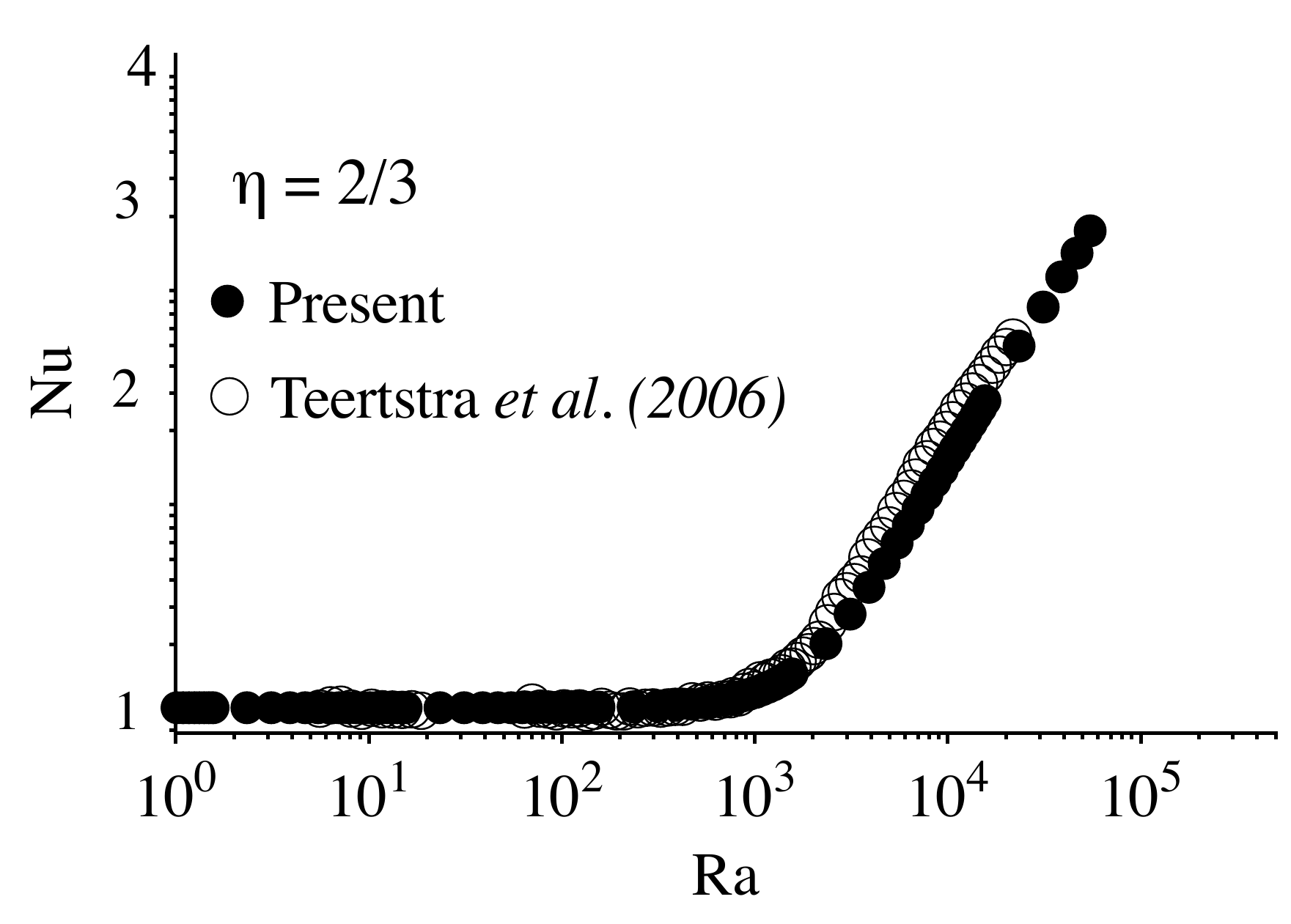}}
    \put(3.2,0){\includegraphics[width=0.35\columnwidth]{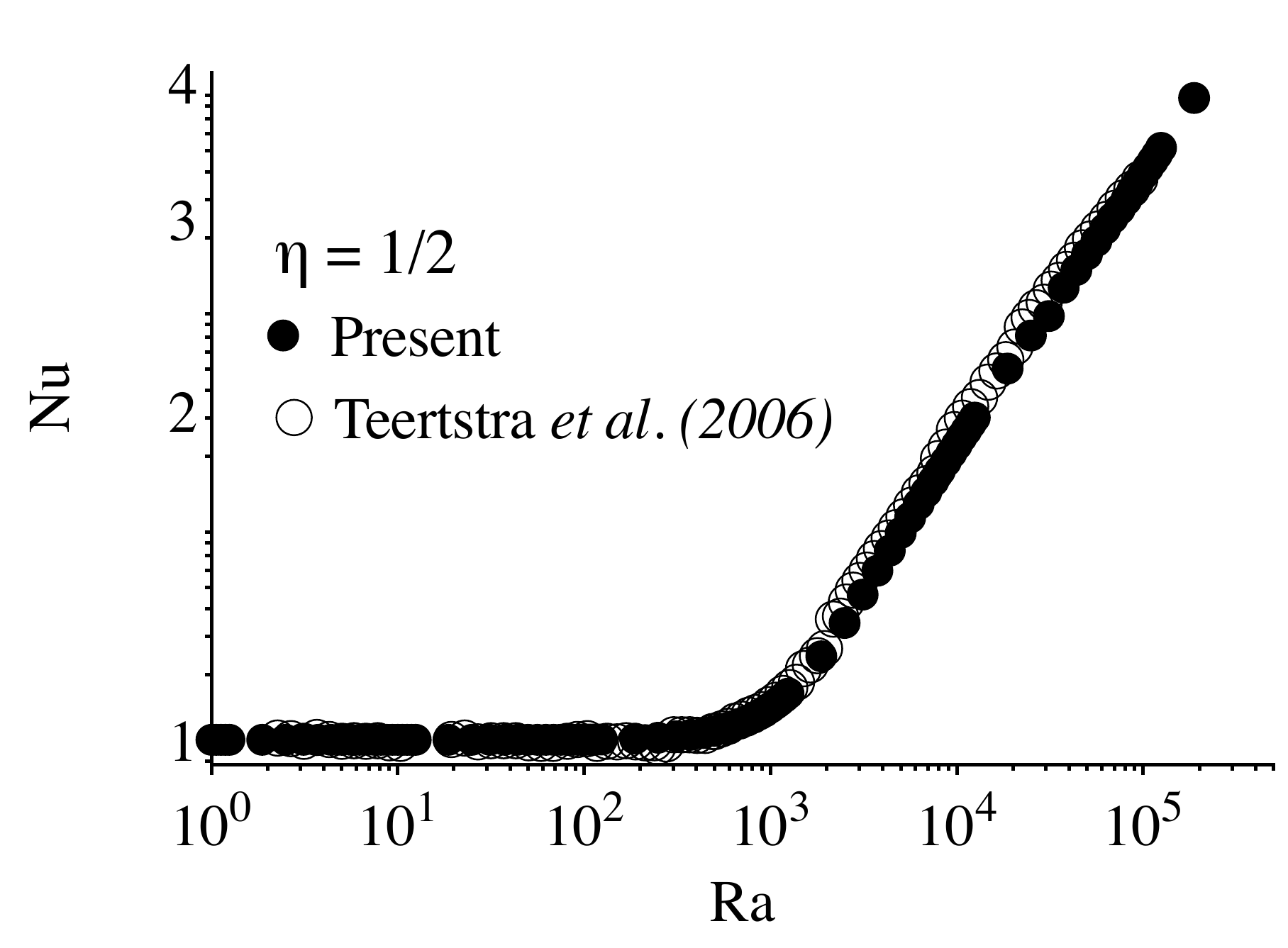}}
    \put(0,0){\includegraphics[width=0.35\columnwidth]{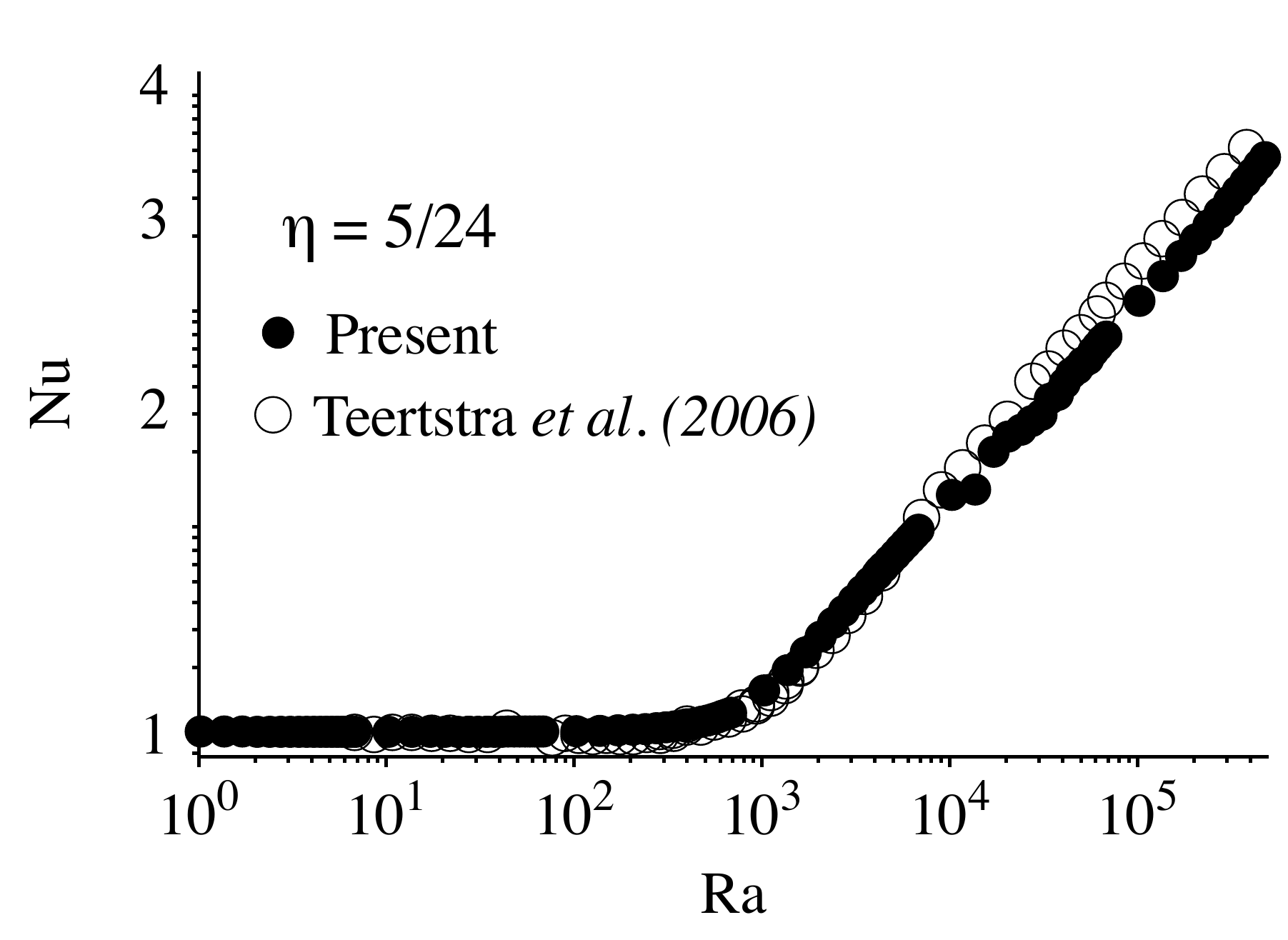}}
    \put(1.5,6.1){(a)}
    \put(1.9,2.3){(b)}
    \put(5.1,2.3){(c)}
    \put(8.3,2.3){(d)}
  \end{picture}
\caption{ Comparison of the steady state $Nu$ numbers obtained with LBM and experiments performed by
 Teertstra {\it et al}. \cite{Teertstra2006} for concentric shell and different aspect ratios. $\eta = 1/3$ in (a), 
 $\eta = 5/24$ in (b), $\eta = 1/2$ in (c) and $\eta = 2/3$ in (d). Insets in (a) show typical isothermal lines in the left half, and steam lines in the right half, found in the three regions divided by the dotted vertical lines.}
\label{fig2}
\end{figure}

To validate the code we compared with a variety of results in the literature. Numerical results reproduced the 
transition from conductive to convective energy transport observed experimentally for concentric spheres by 
Teertstra {\it et al}. \cite{Teertstra2006} through the measurement of the average Nusselt number on the internal 
shell, defined here as
\begin{equation}
 Nu = \frac{L}{4 \pi r_e r_i \Delta T} \oint - \frac{\partial T}{\partial n} dA,
\label{nu}
\end{equation}
\noindent where $n$ is the normal direction to the spheres.

The values of $Nu$ obtained for steady axisymmetric flows are compared to 
those of  Teertstra {\it et al}. in figure \ref{fig2} for different values of $\eta$, showing  a good 
agreement with 
experiments. The insets in figure \ref{fig2}(a) show isothermal curves (left half) and stream lines (right half) 
obtained with the code on a plane of symmetry. Each inset shows the obtained flow patterns and 
plume formation through three characteristic regions, roughly divided by the vertical dotted lines. These regions 
correspond to the conductive and convective regimes, and a transitional region in between them.

A linear fit of the numerical results for the steady convective region in the range $0.2< \eta< 0.7$ suggests a 
power law  relation given by
\begin{equation}
Nu = 0.36\left(1-\eta \right) Ra^{0.23 \eta + 0.14}.
\label{fit}
\end{equation}
This relation is consistent with correlations obtained by Raithby \& Hollands \cite{Raithby1975}, Scanlan {\it et 
al.} \cite{Scanlan1970} and Feldman \& Colonius \cite{Colonius2013}. 

In figure \ref{fig1}(b) isothermal curves and stream lines are shown for a case of vertically eccentric spheres. 
The flow pattern is in good agreement to that observed experimentally by Powe {\it et al.} \cite{Powe1980} for the 
set of parameters in the region they named ``Steady Multiple Interior Cells''. 

\begin{figure}[t]
\centering
\setlength{\unitlength}{0.1\textwidth}
  \begin{picture}(8,5)
    \put(0.5,2.5){\includegraphics[width=0.3\columnwidth]{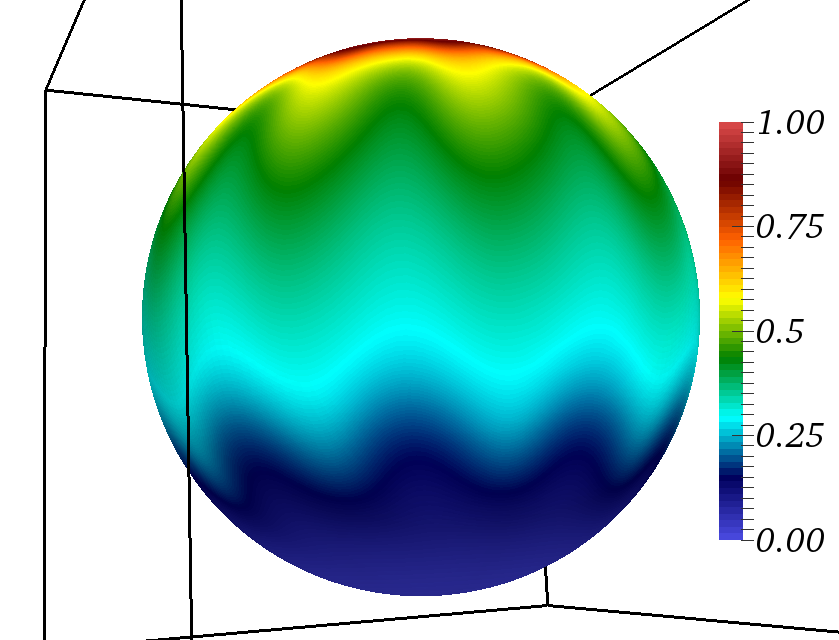}}
    \put(4.5,2.5){\includegraphics[width=0.3\columnwidth]{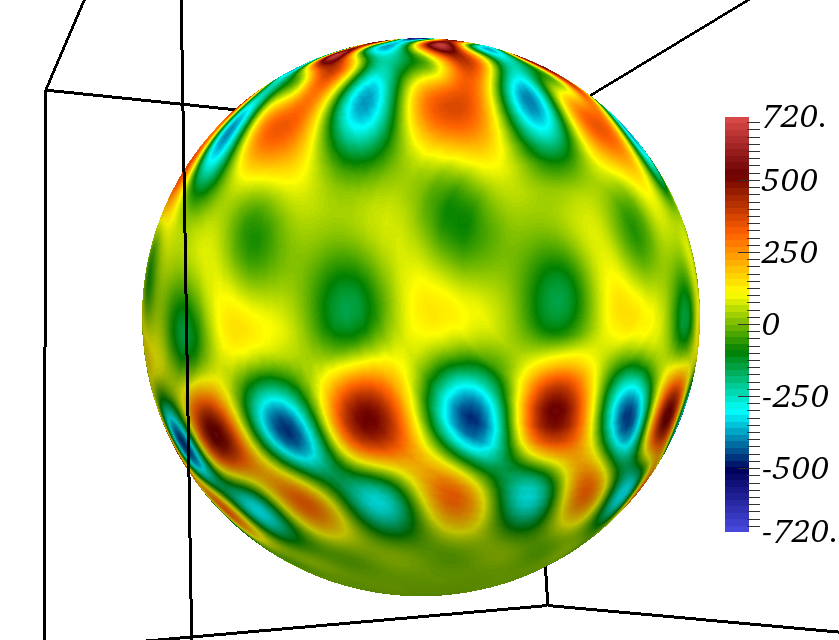}}
    \put(0.5,0){\includegraphics[width=0.3\columnwidth]{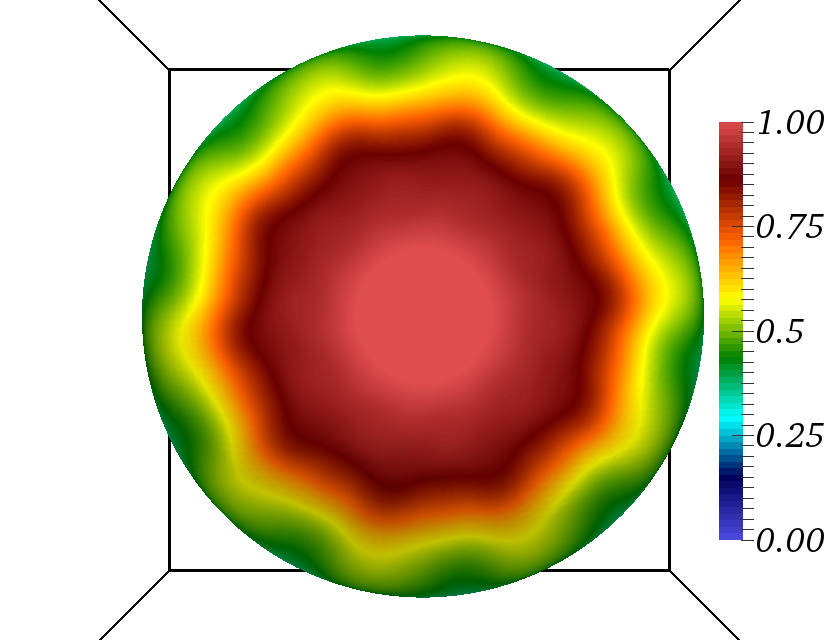}}
    \put(4.5,0){\includegraphics[width=0.3\columnwidth]{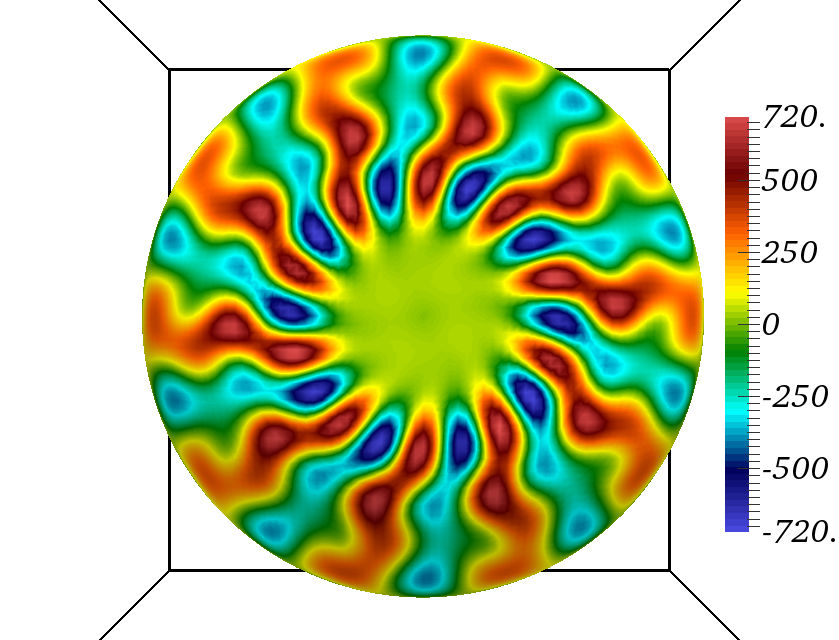}}
    \put(0.4,4.7){(a)}
    \put(4.4,4.7){(b)}
    \put(3.04,4.45){$\Theta$}
     \put(3.04,1.95){$\Theta$}
    \put(7.1,4.5){$u_{\phi}$}
     \put(7.1,2){$u_{\phi}$}
   \put(0.4,2.1){(c)}
    \put(4.4,2.1){(d)}
  \end{picture}
\caption{Numerical solutions of the non-dimensional temperature and azimuthal velocity component for $\eta = 0.714$, $\epsilon = 0$ and $Ra = 4.03 \times 10^4$ on the midrange spherical surface  $r=(r_e + r_i)/2$ showing an azimuthal mode $m=10$. In (a) and (c) $\Theta$ is shown from the top and lateral views respectively. In (b) and (d) $u_{\phi}$ is shown from the top and lateral views respectively.}
\label{fig3}
\end{figure}

\begin{table} [b]
\caption{Steady and time averaged $Nu$ numbers}
\label{Tab1}
\begin{center}
\scalebox{0.8}{
\begin{tabular}{c c c c c c c | c}
\hline \hline 
$\eta$ & Ra & Present & Ref. \cite{Colonius2013} & Ref. \cite{Chu1993} & Ref. \cite{Dehghan2010} 
& Ref. \cite{Chiu1996} & Eq. (\ref{fit})\\ \hline \hline
$0.5$ & $10^2$ & $1.0077$ & $1.0217$ & $1.001$ & $1.000$ &  &\\ 
$0.5$ & $10^3$ & $1.0865$ & $1.104$ & $1.0990$ & $1.1310$ & $1.1021$ & $1.0478$\\ 
$0.5$ & $10^4$ & $1.8977$ & $1.9665$ & $1.9730$ & $1.9495$ & $1.9110$ & $1.88848$\\ 
$0.5$ & $10^5$ & $3.3077$ & $3.4012$ & $3.4890$ & $3.4648$ & $3.355$ & $3.3906$\\ \hline 
$0.667$ & $10^3$ & $1.0415$ & $1.04825$ & $1.001$ & $1.00115$ &  &\\
$0.667$ & $10^4$ & $1.7295$ & $1.793^*$ & $1.073$ & $1.07138$ & & $1.7886$ \\  \hline 
$0.833$ & $10^3$ & $1.0118$ & $1.011$ & $1.0$ & $1.0018$ &  &\\  
$0.833$ & $10^4$ & $1.5993^*$ & $1.6523^*$ & $1.001$ & $1.0028$ & & \\ \hline 
\end{tabular}}
\end{center}
\end{table}

Convection between concentric spheres for $\eta = 0.714$ served to validate the code with unsteady three 
dimensional flows. Figure \ref{fig3} shows numerical solutions for $Ra = 4.03 \times 10^4$, just above the 
critical value ($Ra = 3.96 \times 10^4$) predicted by the linear stability analysis presented by Travnikov {\it et 
al}. \cite{Travnikov2015}. The non-dimensional temperature $\Theta = (T-T_i)/\Delta T$ and azimuthal velocity 
component $u_{\phi}$ on the midrange spherical surface $r=(r_e + r_i)/2$ show an azimuthal mode $m=10$, which is 
also the critical wave number predicted by linear stability. This unsteady periodical flow was obtained by letting 
the solution reach a steady state at $Ra = 3.8 \times 10^4$ before increasing its value to $Ra = 4.03 \times 
10^4$. The pattern of $u_{\phi}$ is very similar to that 
obtained by Scurtu {\it et al.} \cite{Scurtu2010} for travelling and pulsating waves of mode $m=10$ using spectral methods, and 
those obtained by Feldman \& Colonius \cite{Colonius2013} for mode $m=12$ at higher $Ra$ values using openFoam 
algorithms. When $Ra$ was increased to $5 \times 10^4$, the solution showed an $m=11$ mode over many 
oscillations before changing to an $m=12$ mode, which is consistent with results in 
\cite{Scurtu2010,Colonius2013}. The non-dimensional velocity  $u_{\phi}$ shown in figure \ref{fig3}, rescaled 
taking $\alpha = 2 \times 10^{-5} m^2/s$ and $L = 0.1m$, predicts azimuthal velocities of the order of $0.1 m/s$.

In general, steady and time averaged $Nu$ numbers predicted by the code are in good agreement with the 
corresponding values previously reported for steady and periodic flows. This is summarised in Table \ref{Tab1},  
where numbers with an asterisk superscript denote a time average $Nu$ of a periodic solution. Notice that a 
steady state was obtained for $\eta=0.667$ and $Ra = 10^4$, this is in agreement with experiments in 
\cite{Teertstra2006} (see figure \ref{fig2}(c)) .

Typical simulations required $4\times10^5$ time steps and the three dimensional fluid domain was represented 
by as much as $2\times 10^7$ grid points for unsteady solutions. Using a Tesla C2075 parallel processor, runs 
took between three and six days.

\section{Eccentric configurations}

\begin{figure}[t]
\begin{center}
   \includegraphics[width=0.6\columnwidth]{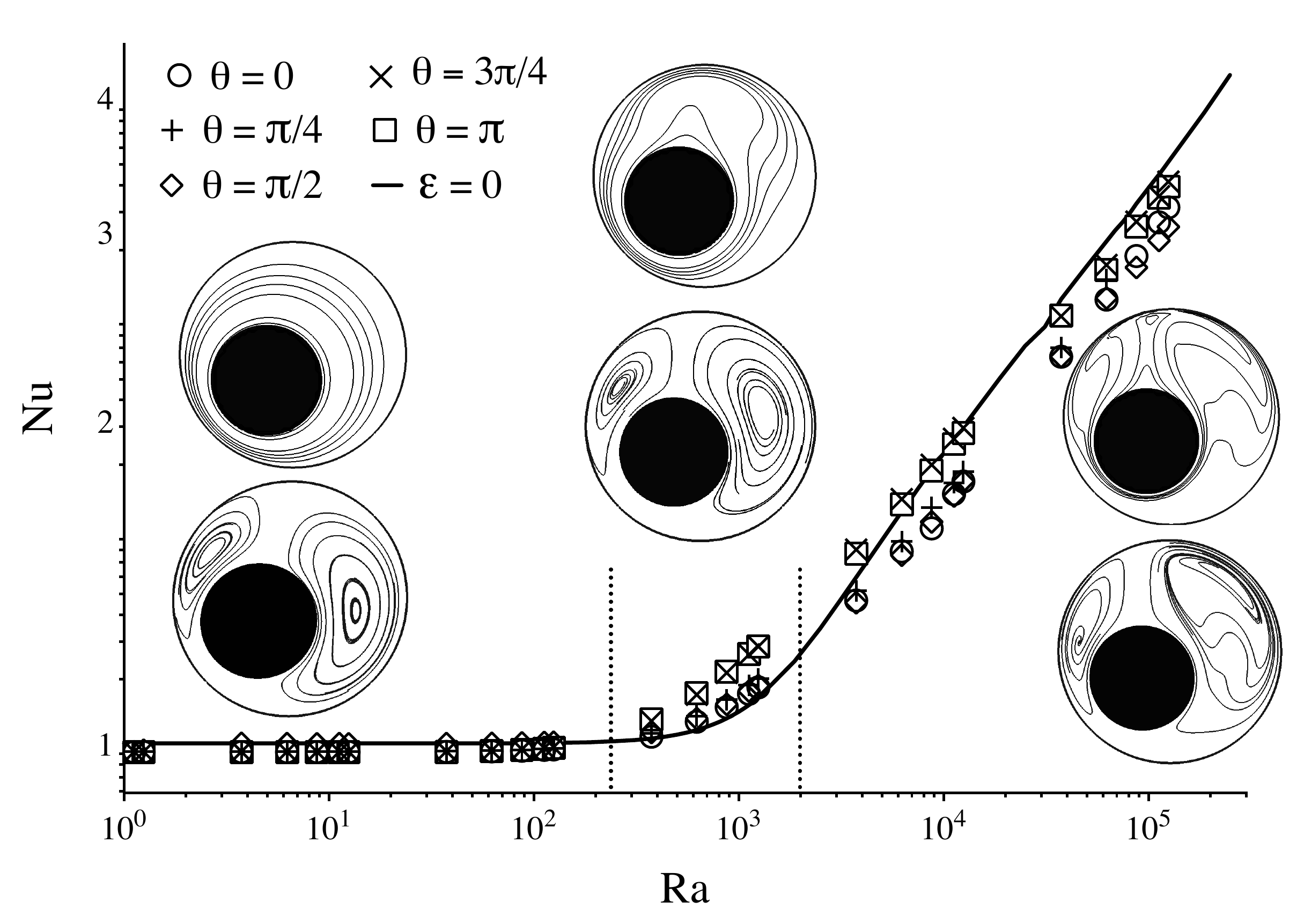}
 \caption{Numerical prediction of the steady $Nu$ number showing the convective to conductive transition for 
 eccentric configurations for $\eta=0.5$ and $\epsilon = 0.6$. Solutions at polar angles $\theta = 0,\pi/4, \pi/2, 3 \pi /4, \pi$ are compared with the concentric solution represented by the continuous line. The insets show the typical isothermal curves (top) and stream lines (bottom) for $\theta = 3 \pi / 4$ on the plane of symmetry in the conductive and convective regions, and a transitional region in between them marked by the vertical dotted lines.}
 \label{fig4}
\end{center}
\end{figure}

For the case of concentric shells, the average $Nu$ number defined in (\ref{nu}) represents the average heat flux 
on either the external or the internal sphere, divided by $ k \Delta T \eta /L$ or $ k \Delta T /( L\eta )$ respectively.
The latter expressions are the average heat flux for the convection problem in concentric spheres in the 
external and internal spheres respectively. Defined in this way $Nu$ tends to one in both spheres for small values 
of $Ra$ as heat transport is dominated by conduction (see figure \ref{fig2}). For eccentric spheres we defined the 
average $Nu$ number accordingly as
\begin{equation}
 Nu = \frac{L}{4 \pi r^2 \Delta T Nu^*} \oint - \frac{\partial T}{\partial r} dA,
\end{equation}
\label{nue}
\noindent where $r$ takes the values $r_e$ and $r_i$, and $ Nu^*$ is the average Nusselt number for the heat 
conduction problem between eccentric spheres obtained by Alassar \cite{Alassar2010}. Through bispherical 
coordinates, $ Nu^*$ is expressed as an infinite series
\begin{equation}
 Nu^* = \frac{2 a L}{r^2 \Delta T} \sum\limits_{n=0}^{\infty} \frac{1}{e^{(2n+1) \xi_e} -  e^{(2n+1) \xi_i} },
\end{equation}
\label{nue}
\noindent where $r$ takes the values $r_e$ and $r_i$, $ \xi_e = {\text sinh}^{-1}(a/r_e)$, $ \xi_i = {\text sinh}^{-1}
(a/r_i)$ and $$a = \sqrt{(c+r_i+r_e)(c+r_i-r_e)(c-r_i+r_e)(c-r_i-r_e)}/2c$$ determines the foci of the bispherical 
coordinates.  

We first searched for steady solutions for fixed values of $\eta$ and $\epsilon$ and different angles $\theta$.
In figure \ref{fig4} appears the behaviour of the steady $Nu$ number as $Ra$ increases for $\eta = 0.5$, $\epsilon 
= 0.6$, and the set of angles $\theta = 0, \pi/4, \pi/2, 3 \pi /4, \pi$. Numerical solutions show a similar 
behaviour of that observed 
in concentric spheres, given by the continuous line in the figure. There is a transition from a conductive to a 
convective regime at roughly the same values 
of the $Ra$ number as the concentric solutions. At angles $\theta = 3 \pi /4$ and $\pi$ the steady convective 
regime seems to be reached at lower values of $Ra$ than the rest of cases shown in the figure. Also, the slopes 
of the convective regime are slightly lower for the eccentric configurations. A linear fit suggest exponents 
of values around $0.245$ for $\theta > \pi/2$, $0.225$ for $\theta < \pi/2$ and its lowest value of 
$0.2$ for $\theta = \pi/2$. Insets show the evolution of 
isothermal curves and steam lines. On the other hand, eccentric simulations become unstable at lower values 
of the $Ra$ number when compared to concentric ones. 

\begin{figure}[t]
\begin{center}
   \includegraphics[width=0.55\columnwidth]{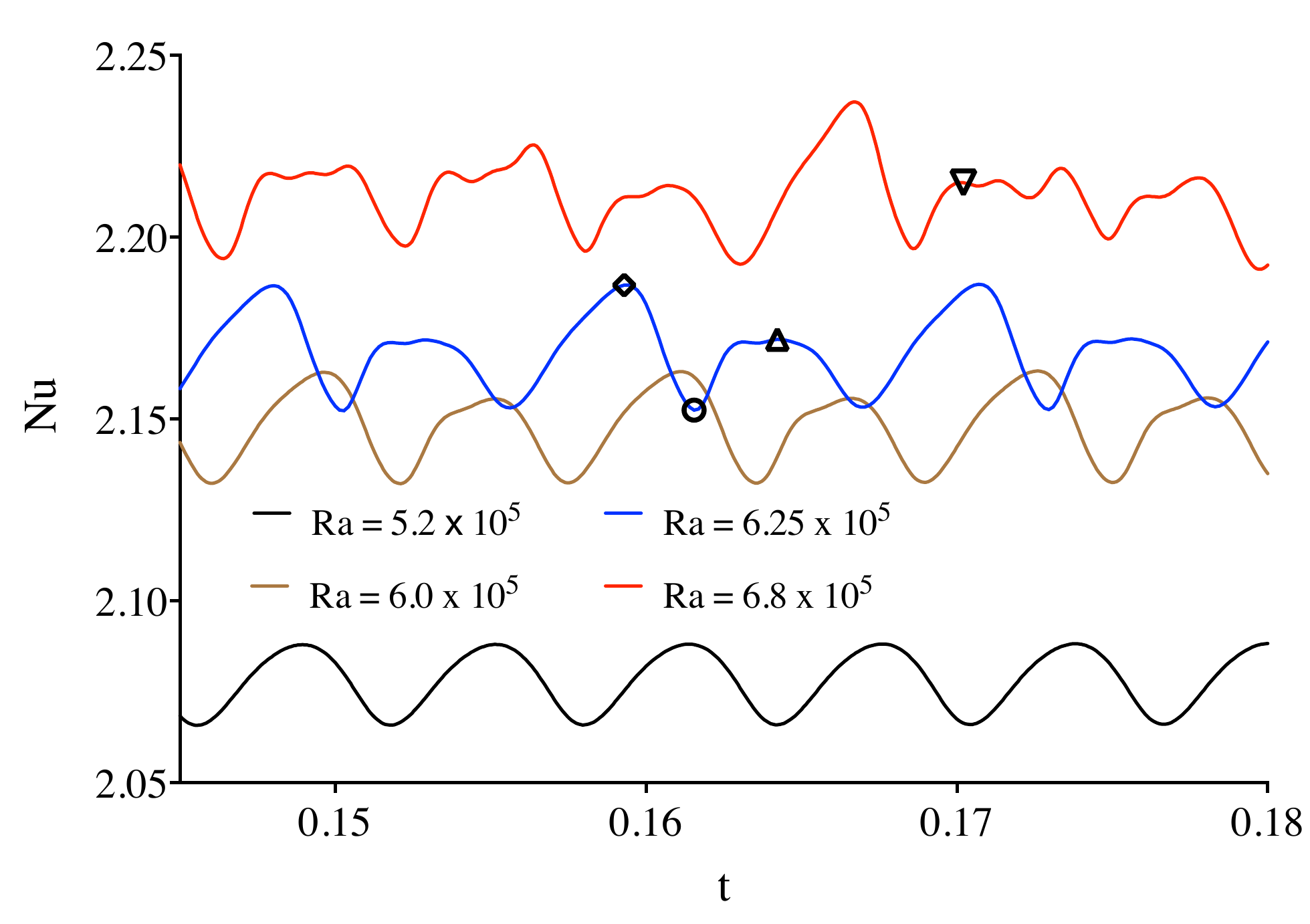}
 \caption{Time evolution of the average $Nu$ number calculated on the external sphere for the case of $\eta= 0.5$, $\epsilon = 0.8$ and $\theta = \pi/4$ and increasing values of the 
 $Ra$ number. The different curves show a periodic behaviour of increasing complexity as the $Ra$ number grows. 
 Details of the flows at the points indicated with the symbols $\diamond$, $\circ$, $\vartriangle$ and $\triangledown$ are shown in figures \ref{fig6} and \ref{fig7}.}
 \label{fig5}
\end{center}
\end{figure}

Beyond the steady convective region a series of periodic solutions where found. Starting from an oscillating 
$Nu$ number at the inner and outer spheres with a clear frequency, as the $Ra$ number 
increased unsteady flow behaviour became irregular through what seamed as a period doubling process. The 
behaviour $Nu$ along non-dimensional time $t$ is shown in figure \ref{fig5} for the case of $\eta= 0.5$, $\epsilon = 0.8$ and $\theta = \pi/4$. Taking $\alpha = 2 \times 10^{-5} m^2/s$ and $L = 0.1m$, the times on figure \ref{fig5} must be rescaled with a time of $500 s$ and so the figure would show a lapse of $10 s$.

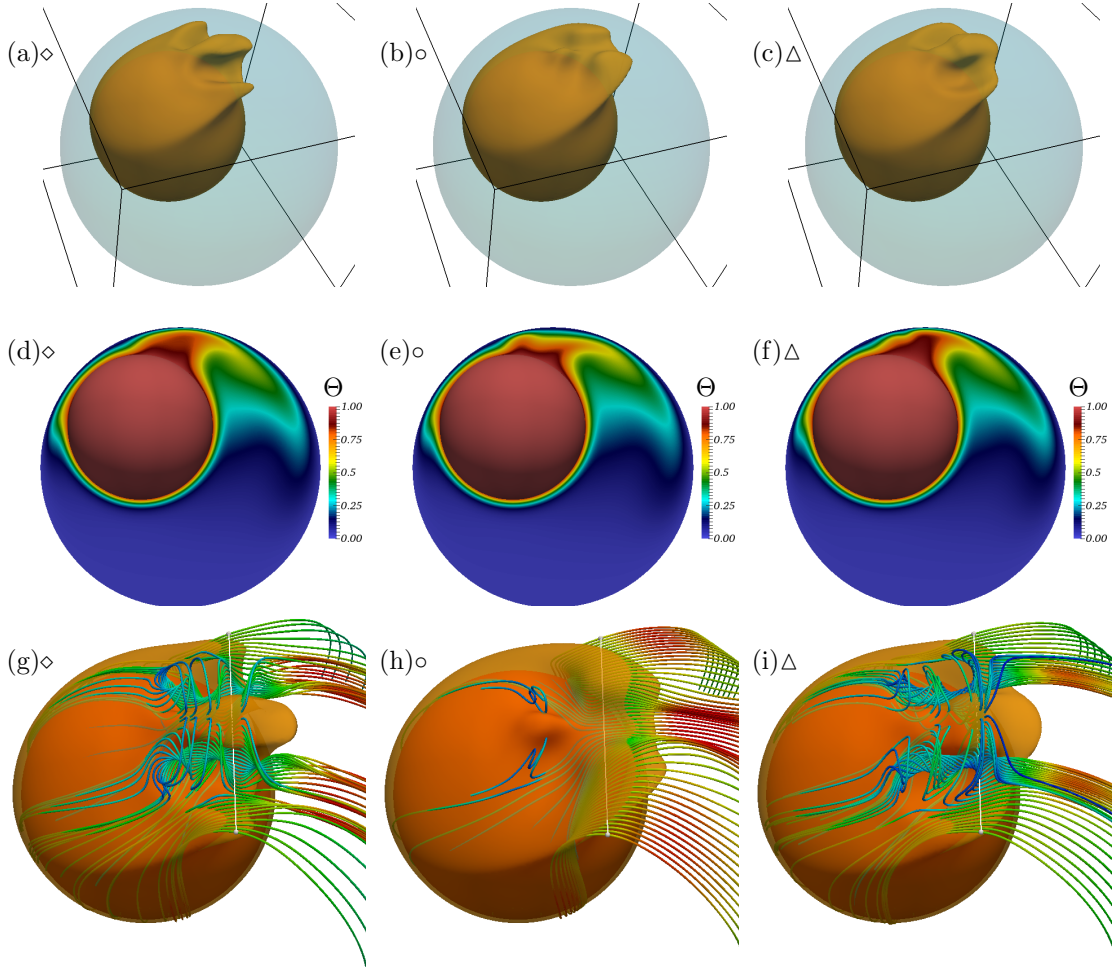
\begin{figure}[t]
\centering
\setlength{\unitlength}{0.1\textwidth}
  \begin{picture}(9,7.5)
    \put(0.4,5.45){\includegraphics[width=0.25\columnwidth]{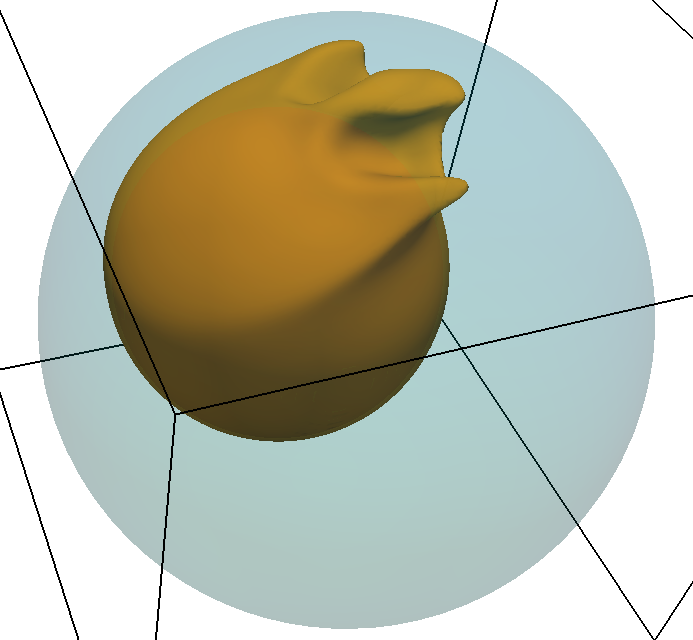}}
    \put(3.4,5.45){\includegraphics[width=0.25\columnwidth]{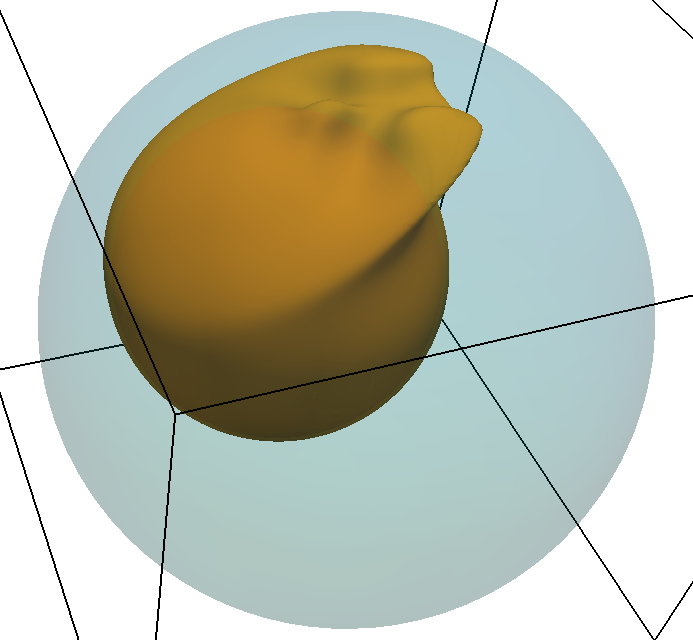}}
    \put(6.4,5.45){\includegraphics[width=0.25\columnwidth]{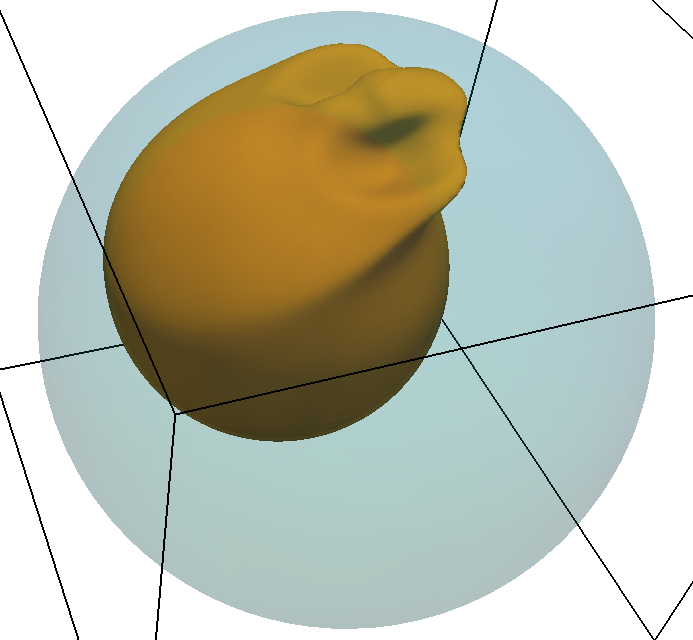}}
    \put(0,2.57){\includegraphics[width=0.3\columnwidth]{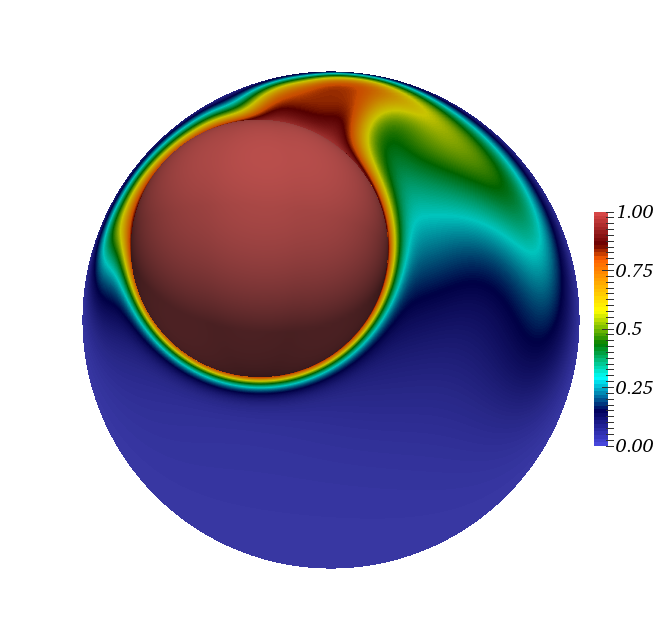}}
    \put(3,2.57){\includegraphics[width=0.3\columnwidth]{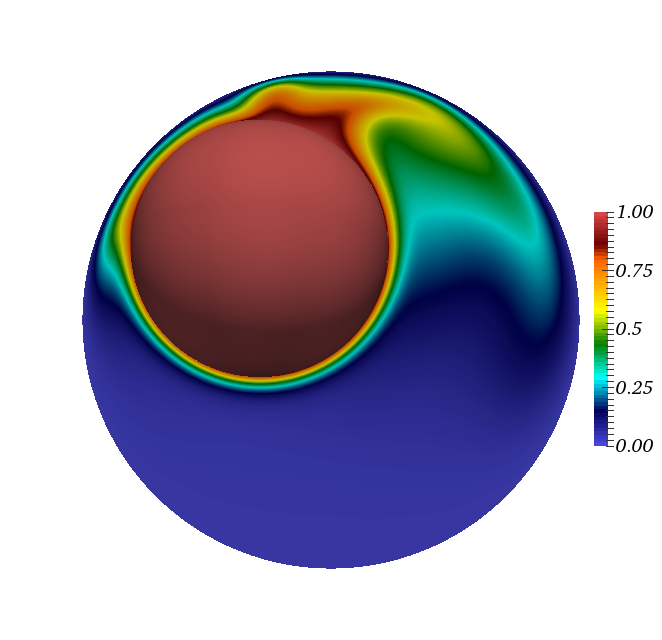}}
    \put(6,2.57){\includegraphics[width=0.3\columnwidth]{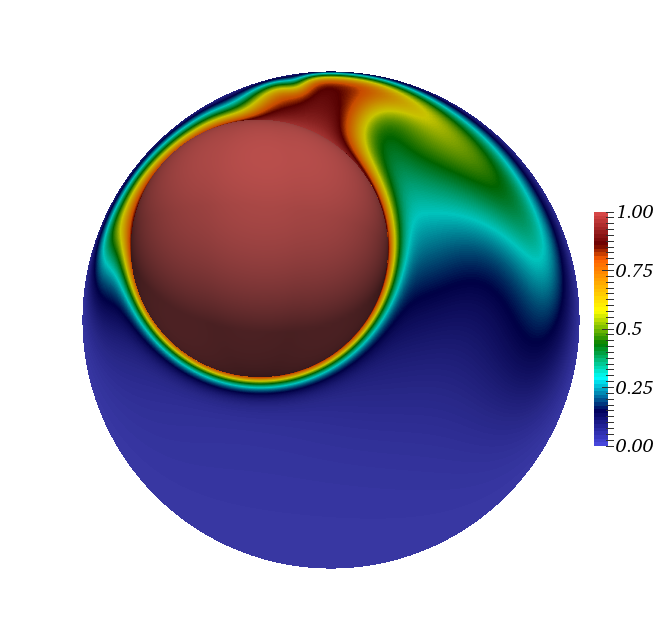}}
    \put(0,0){\includegraphics[width=0.3\columnwidth]{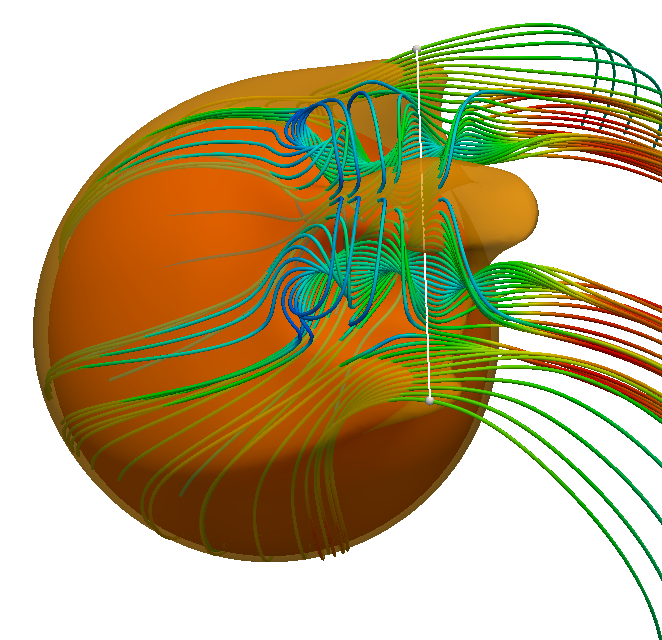}}
    \put(3,0){\includegraphics[width=0.3\columnwidth]{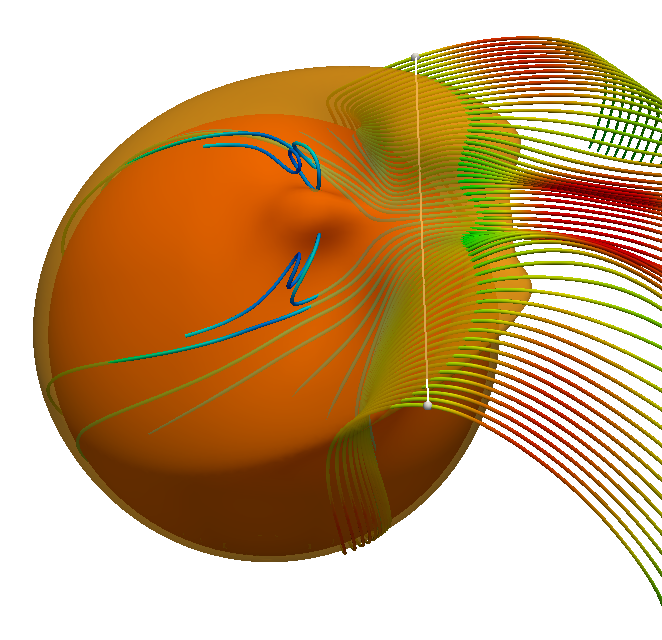}}
    \put(6,0){\includegraphics[width=0.3\columnwidth]{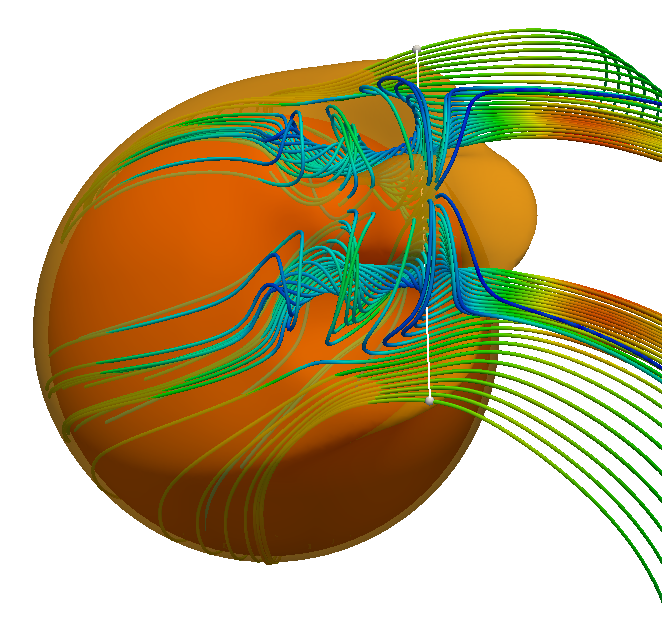}}
    \put(0.1,7.3){(a)$\diamond$}
    \put(3.1,7.3){(b)$\circ$}
    \put(6.1,7.3){(c)$\vartriangle$}
    \put(0.1,4.9){(d)$\diamond$}
    \put(3.1,4.9){(e)$\circ$}
    \put(6.1,4.9){(f)$\vartriangle$}
    \put(0.1,2.4){(g)$\diamond$}
    \put(3.1,2.4){(h)$\circ$}
    \put(6.1,2.4){(i)$\vartriangle$}
    \put(2.65,4.6){$\Theta$}
     \put(5.65,4.6){$\Theta$}
     \put(8.65,4.6){$\Theta$}
  \end{picture}
\caption{Results for the case of $\eta= 0.5$, $\epsilon = 0.8$, $\theta = \pi/4$ and  $Ra=6.8 \times 10^5$ at the times indicated by the points $\diamond$, $\circ$ and $\vartriangle$ in figure \ref{fig5}. The row at the top show surface $\Theta = 0.7$ located inside the enclosing sphere. The row in the middle shows a slice of the temperature field at the plane of symmetry of the configuration. The bottom row shows stream lines seeded on the white line coloured with the magnitude of the velocity.}
\label{fig6}
\end{figure}

Figure \ref{fig6} shows in columns some details of the unsteady convecive flow for $Ra=6.25 \times 10^5$ at   
three times, indicated in figure \ref{fig5}, when the $Nu$ number 
reaches an absolute maximum value, the following minimum and the next local maximum respectively (the symbols 
$\diamond$, $\circ$ and $\vartriangle$ in figure 
\ref{fig5}). The top row shows the isothermal surface $\Theta = 0.7$, the row in the middle shows the temperature 
field on the plane of symmetry of the flow domain and the bottom row shows a set of stream lines superposed to the 
surface of $\Theta = 0.7$. 

\begin{figure}[t]
\begin{center}
   \includegraphics[width=0.4\columnwidth]{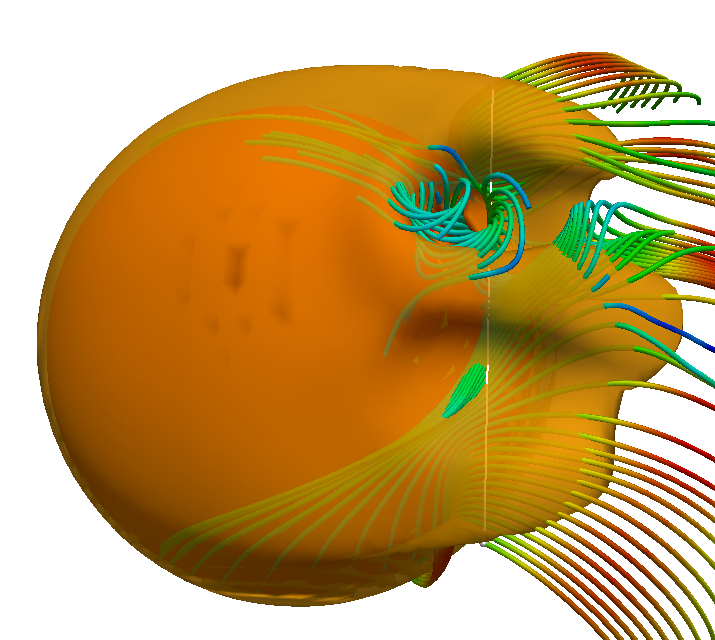}
 \caption{Isothermal surface $\Theta = 0.7$ superimposed to a set of stream lines seeded on the white line at the time indicated by $\triangledown$ in figure \ref{fig5}. Stream lines are coloured with the magnitude of the velocity. Parameter values are  $\eta= 0.5$, $\epsilon = 0.8$, $\theta = \pi/4$ and $Ra=6.8 \times 10^5$. }
 \label{fig7}
\end{center}
\end{figure}

Observations suggest that isothermal surfaces suffer an instability in the direction transversal to the plane of 
symmetry, generating a sort of ``tail'' at the top of the surface that seem to flap vertically. This oscillation 
is accompanied by currents coming from the ``front'' of the inner sphere when the $Nu$ number reaches its peaks 
(see figures \ref{fig6}(g) and \ref{fig6}(i)). Lower values of $Nu$ appear with smoother convection cells and flow 
patterns within the flow (see figure \ref{fig6}(h)). At higher $Ra$ numbers, the currents coming from the front 
start to oscillate horizontally showing a new instability of the isothermal surfaces. This can be seen in figure 
\ref{fig7}, corresponding to the point marked with the symbol $\triangledown$ in the curve of $Ra=6.8 \times 
10^5$ in figure \ref{fig5}.

\section{Conclusions}

It is remarkable that such a simple numerical model as LBM can reproduce steady and unsteady three 
dimensional natural convection flows. Additionally, the simplicity of the algorithm allows for an efficient 
implementation on massively parallel architectures. This was the motivation for exploring natural convection in eccentric annuli configurations 
with this method. Although there is much work done in concentric configurations, eccentric ones have received little 
attention and our observations contribute to the complete picture of natural convection in spherical annuli.
 
When validated with numerical and experimental results of the concentric configuration, results suggested a 
power law correlation of the average $Nu$ number on the spherical boundaries with the $Ra$ number and 
the aspect ratio, see equation (\ref{fit}), close to what has been proposed in previous work (see 
\cite{Raithby1975,Scanlan1970,Colonius2013}).  When simulating eccentric configurations, result suggested a 
transition from a conductive regime to a steady 
convective regime similar to that found in the concentric case. Convective regimes are reached at lower $Ra$ 
numbers and $Nu$ grows at a smaller rate in eccentric configurations. The lack of symmetry should produce 
more mixing in and lower the heat flux observed in the convective regime. 

Eccentric convective states also become unstable at lower values of $Ra$. As $Ra$ grows, flow structure and 
the $Nu$ number present periodic fluctuations in time. $Nu$ oscillations are related to the deformation of 
isothermal surfaces. These show a modulation transversal to the vertical plane of symmetry and, as $Ra$ grows, 
surfaces start to oscillate in the horizontal direction as well. Similar behaviour of the periodicity of the flow and 
heat flux was found in natural convection in a concentric annuli for an inner sphere colder than the outer one 
\cite{Futterer2007}. 

Results here presented are encouraging and suggest that many different examples of three dimensional natural 
convection phenomena between a body and its enclosure can be studied using LBM. Specially given the 
simplicity to impose boundary conditions on curved surfaces in LBM, and the efficiency that exhibits under fine 
grain parallelism.\\
 \\
{\bf Acknowledgements.} 

We are grateful to Dr. P. M. Teertstra for providing experimental data. We are also thankful to Dr. R. Rechtman for 
his fruitful suggestions.

\bibliographystyle{plain}


\end{document}